\documentclass[aps,pre,twocolumn,showpacs]{revtex4}
\usepackage{epsfig}
\usepackage{times}
\begin{document}

\title{Phase transitions induced by variation of invasion rates \\ 
in spatial cyclic predator-prey models with four or six species}

\author{Gy\"orgy Szab\'o and Attila Szolnoki}
\affiliation
{Research Institute for Technical Physics and Materials Science
P.O. Box 49, H-1525 Budapest, Hungary}

\begin{abstract}
Cyclic predator-prey models with four or six species are studied on a square lattice when the invasion rates are varied. It is found that the cyclic invasions maintain a self-organizing pattern as long as the deviation of the invasion rate(s) from a uniform value does not exceed a threshold value. For larger deviations the system exhibits a continuous phase transition into a frozen distribution of odd (or even) label species.     
\end{abstract}


\maketitle

Spatial predator-prey models with cyclic competition exhibit several peculiar features that can be relevant in biological, chemical and ecological systems
\cite{hofbauer_98,durrett_tpb98,johnson_prslb02,szabo_pr07,traulsen_pre04}. In these systems species are distributed on the sites of a lattice and the predators' offspring can be substituted for the neighboring prey. More precisely, for the simplest cyclic $n$-species systems ($n \geq 3$) a randomly chosen individual of species $i$ can occupy the neighboring site of species $i+1 \; \mbox{mod}\; n$ ($i=0, 1, \ldots , n-1$) with an invasion rate independent of $i$. On a one-dimensional lattice ($d=1$) this evolutionary process yields growing domains (of sites occupied by the same species) if $n < 5$; otherwise the system develops into a frozen pattern \cite{frachebourg_pre96}. For higher-dimensional lattices the uniform invasion rate maintains a self-organizing pattern if the system size $N$ goes to infinity and the number of species does not exceed a threshold value dependent on $d$, that is, $n < n_{c}(d)$ [e.g., $n_c(d=2)=14$] \cite{frachebourg_jpa98}. In this case the self-organization is maintained by the unceasing motion of interfaces separating a predator from its prey and the species are present with the same average density ($\rho_i=1/n$) independently of the initial state containing all species.

The mean-field analysis of these systems indicates different types of solutions \cite{hofbauer_98}. Besides the above mentioned symmetric solution ($\rho_i=1/n$) these models have $n$ homogeneous solutions (e.g., $\rho_0=1$ and $\rho_1= \ldots =\rho_{n-1}=0$) which are unstable against the invasion of the corresponding predator. At the same time this system exhibits a continuous set of oscillatory solutions satisfying  conservation laws, e.g., $\sum_i \rho_i=1$ and $\prod_i \rho_i=\mbox{constant}$ \cite{hofbauer_98}. The pair approximation, as a first step to extend the classical mean-field theory \cite{marro_99,szabo_pr07}, confirms the existence of both the symmetric and the $n$ homogeneous solutions. Instead of the periodic oscillations, however, the pair approximation predicts growing oscillations ending in one of the homogeneous states due to the numerical rounding or noise. For the simplest case, called
the evolutionary rock-scissors-paper game ($n=3$), the latter behavior was observed on Bethe lattices of degree $z \ge 6$ in the limit $N \to \infty$ \cite{szabo_jpa04,sato_mmit97} (while a limit cycle is reported for $z=3$ or 4). The real behavior, observed by means of Monte Carlo (MC) simulations, on a square lattice (for $n=3$) could be reproduced by the more sophisticated four-site approximation determining the probability of each possible configuration on the $2 \times 2$ cluster of sites \cite{szabo_jpa04,szabo_pr07}. The failure of the pair approximation for spatial models indicates that their behavior is sensitive to the details of the lattice structure. Similar relevant sensitivity is found when considering the effect of different (species-specific) invasion rates.

When the simplest ($n=3$) case is considered, the variation of invasion rates modifies the composition of the central (stationary) solution \cite{hofbauer_98}. Paradoxically, if one of the species is fed more effectively by increase of its invasion rate then its predator will benefit from this modification \cite{tainaka_pla93,frean_prsb01}. In other words, if the largest invasion rate is from species 0 to 1 then species 2 (the predator of 0) will have the highest density in the central stationary state. This mean-field prediction is confirmed by MC simulations on two-dimensional lattices. The unusual response of the system is related to the indirect interaction mediated through the food web. For example, if the density of species 0 is increased by any external effect then it yields a decrease in the density of species 1, which consume less of species 2, and so on. For a cyclic food web with odd number of species this mechanism results in a negative feedback, in agreement with the mentioned results in the stationary state. For even number of species, however, the feedback becomes positive and may yield the extinction of the odd (or even) label species. 

Using mean-field analysis, pair approximation, and MC simulations, 
this parity effect was demonstrated by Sato {\it et al.} \cite{sato_amc02} who considered a cyclic predator-prey model where each invasion rate is chosen to be unity excepting the rate ($\alpha$) from species 0 to 1. According to their results for even $n$ the odd label species die out if $\alpha < 1$, while the even label species become extinct if $\alpha > 1$. Here $\alpha = 1$ denotes a singular point where the symmetric solution ($\rho_i = 1/n$) remains stable. Unfortunately, the authors did not study the close vicinity of this singular point in their MC simulations.  

Now we show that, instead of sudden changes in $\rho_i$ at $\alpha =1$, the systems exhibit two consecutive phase transitions referring to the existence of an $n$-species state with continuously varying composition. We also study the critical behavior of the phase transition separating the frozen and coexisting states.

\begin{figure}[ht]
\centerline{\epsfig{file=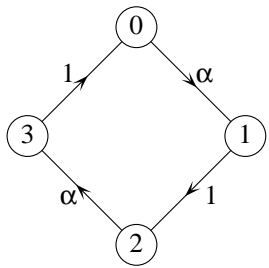,width=4cm}}
\caption{\label{fig:foodweb4} Invasion rates between the predator-prey pairs for the four-species cyclic food web.}
\end{figure}

In order to improve accuracy, first we consider the symmetric version of the model suggested by Sato {\it et al}. \cite{sato_amc02} for $n=4$. It should be mentioned that the qualitative results obtained for the symmetric model agree with those valid for the original model as discussed later. We assume that each site $x$ of a square lattice is occupied by a single individual belonging to one of the $n$ species and their spatial distribution is described by a set of site variables $s_x=i$ where $i=0, \ldots, (n-1)$ refers to the label of the resident species. The system evolution is governed by the repetition of invasion between two neighboring sites chosen at random 
if the sites are occupied by a predator-prey pair [i.e., $(s_x,s_y) \to (s_x,s_x)$ if $s_x$ is the predator of $s_y$]. The species-specific invasion probabilities are proportional to the invasion rates denoted in Fig.~\ref{fig:foodweb4} and backward invasions are forbidden. Notice that if a suitable new time scale is chosen for $\alpha > 1$ this system becomes equivalent to a model with $\alpha <1$ if the species labels are increased by 1.

MC simulations are performed for different values of $\alpha$ on a square lattice with periodic boundaryy conditions and the linear size $L$ of the system is varied from 600 to 4000. The large system size is chosen to avoid artifact effects like trapping to an absorbing state as a consequence of small system size. During one time unit, called a MC step (MCS), each site has a chance once on average to be invaded and to invade one of the neighboring sites. The system is started from a random initial state and after a suitable thermalization time $t_t$ the average density of species ($\rho_i$) is determined by averaging over a sampling time ($t_s$). The thermalization and sampling times ($t_t \simeq t_s \agt 5 \tau_r$ where $\tau_r$ is the relaxation time) are varied from 10 000 to 200 000 MCSs. Clearly, the largest values of $L$, $t_t$, and $t_s$ are used in the close vicinity of the transition points where the relaxation time, correlation length, and fluctuations diverge algebraically. Due to the translational symmetry in the invasion rates, the odd (even) label species have the same density, that is, $\rho_0=\rho_2$ and $\rho_1=\rho_3=1/2-\rho_0$. Henceforth our analysis will be focused on the limit $N \to \infty$.

\begin{figure}[ht]
\centerline{\epsfig{file=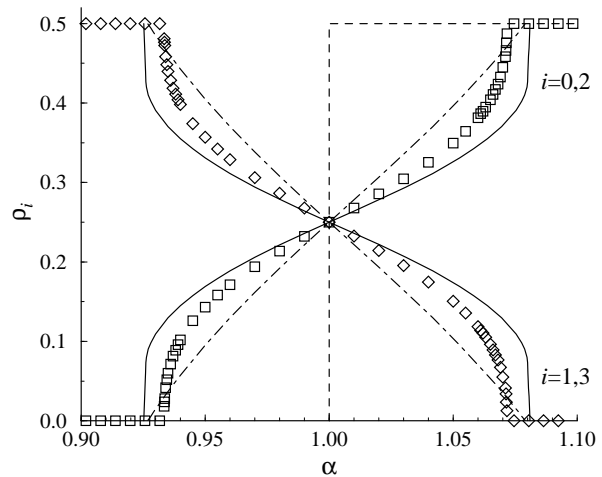,width=8cm}}
\caption{\label{fig:rhoa1a1}Average density of species as a function of $\alpha$ characterizing the invasion rates from the even label species as denoted in Fig.~\ref{fig:foodweb4}. Diamonds and squares indicate MC results while the dashed line shows the prediction of the pair approximation \cite{sato_amc02} for the even label species. The results of $2 \times 2$- and $3 \times 3$-site approximations are illustrated by dash-dotted and solid lines. }
\end{figure}

According to the MC results (see Fig.~\ref{fig:rhoa1a1}), two consecutive critical transitions can be observed for $\alpha_{c_{1}}=0.9332(1)$ and $\alpha_{c_{2}}=1/\alpha_{c_{1}}$. For $\alpha <\alpha_{c1}$ the simulations end in a frozen spatial distribution of species 1 and 3. Clearly, the even label species (0 and 2) form a frozen pattern in the final state if $\alpha > \alpha_{c_{2}}$. Both types of frozen pattern can be considered as an infinitely degenerate absorbing state from which the system cannot escape. Within the intermediate region the four species coexist and invade each other cyclically. In this phase, the average density of species varies monotonically with $\alpha$ as illustrated in Fig.~\ref{fig:rhoa1a1}. This feature is well reproduced by generalized mean-field techniques where we have determined the probability of each possible configuration on 
$2 \times 2$- or $3 \times 3$-site clusters on a square lattice (for a brief description of this method see Appendix C in Ref.~\cite{szabo_pr07}). Clearly, the analytical approximations predict mean-field type transitions and the numerical values for the first critical point are: $\alpha_{c_{1}}^{(4p)}=0.9267(2)$ and $\alpha_{c_{1}}^{(9p)}=0.9252(1)$. 

\begin{figure}[ht]
\centerline{\epsfig{file=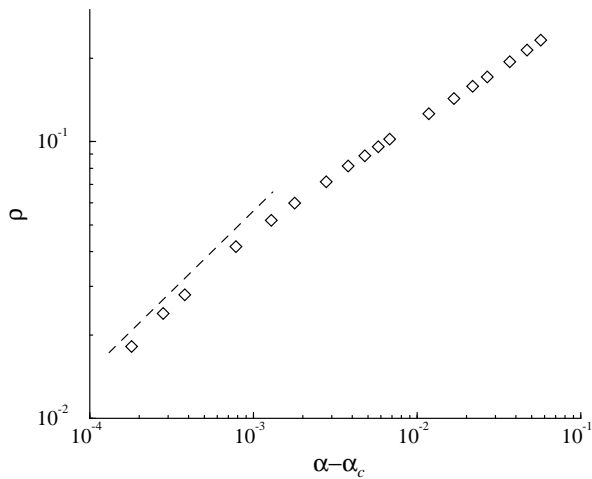,width=8cm}}
\caption{\label{fig:critexp}Log-log plot of the vanishing density of species vs. $\alpha - \alpha_{c_{1}}$. The dashed line with a slope of 0.58 indicates the power law behavior characteristic of the directed percolation universality class in two-dimensional systems.}
\end{figure}

The rigorous analysis of the MC data in the vicinity of the first transition point indicates critical behavior. Namely, the density of the vanishing species, as an order parameter, follows a power law behavior when approaching the critical point, that is, $\rho_0=\rho_2 \simeq (\alpha - \alpha_{c_{1}})^{\beta}$ where $\beta =0.58(1)$ as illustrated in Fig.~\ref{fig:critexp}. Similar behavior characterizes systems belonging to the directed percolation (DP) universality class. In fact, when the transition point is approached, the extinction of the four-species cyclic phase for sufficiently large time and length scales becomes analogous to the elimination of infected sites in the contact process \cite{liggett_85}.  

The universal behavior of the directed percolation universality class \cite{grassberger_zpb82} occurs typically on homogeneous backgrounds \cite{marro_99}. In the present case, however, the extinction process leaves an inhomogeneous, frozen pattern behind. Consequently this system has infinitely many absorbing states. Considering a much simpler system, Jensen \cite{jensen_prl93} has demonstrated that in such systems the non-equilibrium phase transition may also belong to the directed percolation universality class. 

\begin{figure}[ht]
\centerline{\epsfig{file=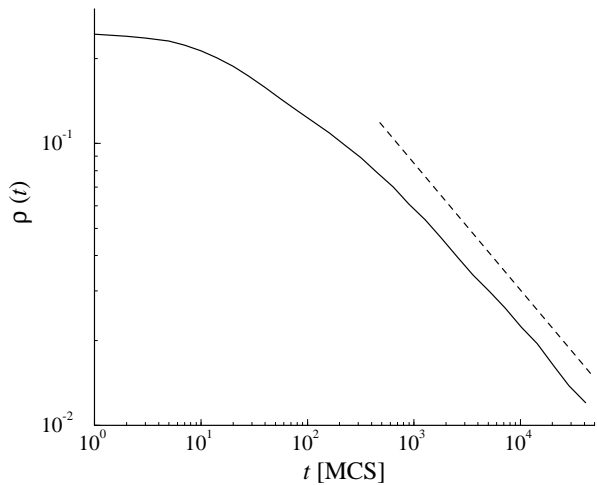,width=8cm}}
\caption{\label{fig:rho_t}Log-log plot for the time dependence of the density of the vanishing species at the critical point ($L=3000$). Fluctuations in $\rho_0(t)=\rho_2(t)$ are suppressed by averaging over a time window with a typical width of $t/6$. The dashed line (with a slope of -0.45) illustrates the power law behavior characteristic of the directed percolation universality class.}
\end{figure}

The present MC results give another example where the universal behavior is not affected by some types of inhomogeneities. In other words, the extinction process is controlled by a background that can be considered homogeneous on sufficiently large time and length scales. To confirm this statement, we have also determined the time dependence of the density of the vanishing species at the first critical point. It is well known that for the two-dimensional, directed percolation universality class the order parameter vanishes algebraically, that is, $\rho_0(t)=\rho_2(t) \simeq t^{- \theta}$ where $\theta = 0.45(1)$ \cite{hinrichsen_ap00}. The MC data (see Fig.~\ref{fig:rho_t}) justify this expectation.

The existence of the four-species phase (and phase transitions) is independent of the symmetrical invasion rates illustrated in Fig.~\ref{fig:foodweb4}. According to the MC simulations the four-species phase is also observed for those models where only one invasion rate differs from 1. Following the notation of Sato {\it et al.} \cite{sato_amc02}, we have studied the system where the invasion rate from species 0 to 1 (denoted by $\alpha$) is varied while the other rates remain unity. Here, in the absence of the previously mentioned symmetry all the species densities become different within the four-species phase existing if $\alpha_{c_{1}} < \alpha < \alpha_{c_{2}}$ [here $\alpha_{c_{1}}=0.8716(1)$ and $\alpha_{c_{2}}=1.145(1)$]. When $\alpha_{c_{1}}$ is approached from above, the densities of even label species decrease simultaneously, i.e., $\rho_0, \rho_2 \simeq (\alpha - \alpha_{c_{1}})^{\beta}$ ($\rho_0 > \rho_2$), suggesting that the critical behavior belongs to the DP universality class too. An asymmetric composition can also be observed in two-strategy frozen patterns, namely, $\rho_1 > \rho_3$ if $\alpha < \alpha_{c_{1}}$, and $\rho_2 > \rho_0$ if $\alpha > \alpha_{c_{2}}$. Similar behavior is expected when all the invasion rates are tuned. The robustness of this universal behavior was confirmed for those systems where a slight mixing is introduced by allowing site exchange with a low probability between neighboring neutral species (e.g., 0 and 2 or 1 and 3) \cite{szabo_jpa05,szabo_pre04a}. 

In addition to the four-species systems, we have also studied what happens in a six-species cyclic predator-prey model if the invasion rate $\alpha$ from species 0 to 1 is varied while the others are chosen to be unity. In the spirit of the parity law \cite{sato_amc02}, the cyclic food web mediates an indirect effect yielding the behavior described above. MC simulations have confirmed that this system evolves from the random initial state to a frozen pattern consisting of odd (even) label species if $\alpha < \alpha_{c_{1}}=0.8971(3)$ [$\alpha > \alpha_{c_{2}}=1.1196(3)$]. Within the intermediate phase ($\alpha_{c_{1}} < \alpha < \alpha_{c_{2}}$) all six species survive and their densities vary similarly to those plotted in Fig.~\ref{fig:rhoa1a1}. According to our previous paper \cite{szabo_pre07}, similar behavior is expected for $n=6$ if some weak mixing is allowed, while the unusual behavior within a range of the strength of mixing for $n=8$ and $10$ raises further questions. 

In summary, the present investigation revises the parity law deduced originally from the results of the pair approximation for spatial cyclic predator-prey models with even number of species. The more sophisticated levels of the dynamical mean-field approximation and also the MC simulations show that a (noisy) cyclic invasion is capable of maintaining a state in which all the species survive with a composition depending on the invasion rates if the unilateral deviation from the uniform rate does not exceed a threshold value. When the invasion rate(s) is varied the system exhibits (directed percolation type) critical phase transitions into frozen states where only the odd (or even) label species are present.

\begin{acknowledgments}

This work was supported by the Hungarian National Research Fund
(Grant No. T-47003).

\end{acknowledgments}

%\bibliography{egg}

\begin{thebibliography}{20}
\expandafter\ifx\csname natexlab\endcsname\relax\def\natexlab#1{#1}\fi
\expandafter\ifx\csname bibnamefont\endcsname\relax
  \def\bibnamefont#1{#1}\fi
\expandafter\ifx\csname bibfnamefont\endcsname\relax
  \def\bibfnamefont#1{#1}\fi
\expandafter\ifx\csname citenamefont\endcsname\relax
  \def\citenamefont#1{#1}\fi
\expandafter\ifx\csname url\endcsname\relax
  \def\url#1{\texttt{#1}}\fi
\expandafter\ifx\csname urlprefix\endcsname\relax\def\urlprefix{URL }\fi
\providecommand{\bibinfo}[2]{#2}
\providecommand{\eprint}[2][]{\url{#2}}

\bibitem[{\citenamefont{Durrett and Levin}(1998)}]{durrett_tpb98}
\bibinfo{author}{\bibfnamefont{R.}~\bibnamefont{Durrett}} \bibnamefont{and}
  \bibinfo{author}{\bibfnamefont{S.}~\bibnamefont{Levin}},
  \bibinfo{journal}{Theor. Pop. Biol.} \textbf{\bibinfo{volume}{53}},
  \bibinfo{pages}{30} (\bibinfo{year}{1998}).

\bibitem[{\citenamefont{Johnson and Seinen}(2002)}]{johnson_prslb02}
\bibinfo{author}{\bibfnamefont{C.~R.} \bibnamefont{Johnson}} \bibnamefont{and}
  \bibinfo{author}{\bibfnamefont{I.}~\bibnamefont{Seinen}},
  \bibinfo{journal}{Proc. Roy. Soc. Lond. B} \textbf{\bibinfo{volume}{269}},
  \bibinfo{pages}{655} (\bibinfo{year}{2002}).

\bibitem[{\citenamefont{Szab{\'o} and F{\'a}th}(2007)}]{szabo_pr07}
\bibinfo{author}{\bibfnamefont{G.}~\bibnamefont{Szab{\'o}}} \bibnamefont{and}
  \bibinfo{author}{\bibfnamefont{G.}~\bibnamefont{F{\'a}th}},
  \bibinfo{journal}{Phys. Rep.} \textbf{\bibinfo{volume}{446}},
  \bibinfo{pages}{97} (\bibinfo{year}{2007}).

\bibitem[{\citenamefont{Hofbauer and Sigmund}(1998)}]{hofbauer_98}
\bibinfo{author}{\bibfnamefont{J.}~\bibnamefont{Hofbauer}} \bibnamefont{and}
  \bibinfo{author}{\bibfnamefont{K.}~\bibnamefont{Sigmund}},
  \emph{\bibinfo{title}{Evolutionary Games and Population Dynamics}}
  (\bibinfo{publisher}{Cambridge University Press},
  \bibinfo{address}{Cambridge}, \bibinfo{year}{1998}).

\bibitem[{\citenamefont{Traulsen and Claussen}(2004)}]{traulsen_pre04}
\bibinfo{author}{\bibfnamefont{A.}~\bibnamefont{Traulsen}} \bibnamefont{and}
  \bibinfo{author}{\bibfnamefont{J.~C.} \bibnamefont{Claussen}},
  \bibinfo{journal}{Phys. Rev. E} \textbf{\bibinfo{volume}{70}},
  \bibinfo{pages}{046128} (\bibinfo{year}{2004}).

\bibitem[{\citenamefont{Frachebourg et~al.}(1996)\citenamefont{Frachebourg,
  Krapivsky, and Ben-Naim}}]{frachebourg_pre96}
\bibinfo{author}{\bibfnamefont{L.}~\bibnamefont{Frachebourg}},
  \bibinfo{author}{\bibfnamefont{P.~L.} \bibnamefont{Krapivsky}},
  \bibnamefont{and} \bibinfo{author}{\bibfnamefont{E.}~\bibnamefont{Ben-Naim}},
  \bibinfo{journal}{Phys. Rev. E} \textbf{\bibinfo{volume}{54}},
  \bibinfo{pages}{6186} (\bibinfo{year}{1996}).

\bibitem[{\citenamefont{Frachebourg and Krapivsky}(1998)}]{frachebourg_jpa98}
\bibinfo{author}{\bibfnamefont{L.}~\bibnamefont{Frachebourg}} \bibnamefont{and}
  \bibinfo{author}{\bibfnamefont{P.~L.} \bibnamefont{Krapivsky}},
  \bibinfo{journal}{J. Phys. A} \textbf{\bibinfo{volume}{31}},
  \bibinfo{pages}{L287} (\bibinfo{year}{1998}).

\bibitem[{\citenamefont{Marro and Dickman}(1999)}]{marro_99}
\bibinfo{author}{\bibfnamefont{J.}~\bibnamefont{Marro}} \bibnamefont{and}
  \bibinfo{author}{\bibfnamefont{R.}~\bibnamefont{Dickman}},
  \emph{\bibinfo{title}{Nonequilibrium Phase Transitions in Lattice Models}}
  (\bibinfo{publisher}{Cambridge University Press},
  \bibinfo{address}{Cambridge}, \bibinfo{year}{1999}).

\bibitem[{\citenamefont{Szab{\'o} et~al.}(2004)\citenamefont{Szab{\'o},
  Szolnoki, and Izs{\'a}k}}]{szabo_jpa04}
\bibinfo{author}{\bibfnamefont{G.}~\bibnamefont{Szab{\'o}}},
  \bibinfo{author}{\bibfnamefont{A.}~\bibnamefont{Szolnoki}}, \bibnamefont{and}
  \bibinfo{author}{\bibfnamefont{R.}~\bibnamefont{Izs{\'a}k}},
  \bibinfo{journal}{J. Phys. A: Math. Gen.} \textbf{\bibinfo{volume}{37}},
  \bibinfo{pages}{2599} (\bibinfo{year}{2004}).

\bibitem[{\citenamefont{Sato et~al.}(1997)\citenamefont{Sato, Konno, and
  Yamaguchi}}]{sato_mmit97}
\bibinfo{author}{\bibfnamefont{K.}~\bibnamefont{Sato}},
  \bibinfo{author}{\bibfnamefont{N.}~\bibnamefont{Konno}}, \bibnamefont{and}
  \bibinfo{author}{\bibfnamefont{T.}~\bibnamefont{Yamaguchi}},
  \bibinfo{journal}{Mem. Muroran Inst. Tech.} \textbf{\bibinfo{volume}{47}},
  \bibinfo{pages}{109} (\bibinfo{year}{1997}).

\bibitem[{\citenamefont{Tainaka}(1993)}]{tainaka_pla93}
\bibinfo{author}{\bibfnamefont{K.}~\bibnamefont{Tainaka}},
  \bibinfo{journal}{Phys. Lett. A} \textbf{\bibinfo{volume}{176}},
  \bibinfo{pages}{303} (\bibinfo{year}{1993}).

\bibitem[{\citenamefont{Frean and Abraham}(2001)}]{frean_prsb01}
\bibinfo{author}{\bibfnamefont{M.}~\bibnamefont{Frean}} \bibnamefont{and}
  \bibinfo{author}{\bibfnamefont{E.~D.} \bibnamefont{Abraham}},
  \bibinfo{journal}{Proc. R. Soc. Lond. B} \textbf{\bibinfo{volume}{268}},
  \bibinfo{pages}{1323} (\bibinfo{year}{2001}).

\bibitem[{\citenamefont{Sato et~al.}(2002)\citenamefont{Sato, Yoshida, and
  Konno}}]{sato_amc02}
\bibinfo{author}{\bibfnamefont{K.}~\bibnamefont{Sato}},
  \bibinfo{author}{\bibfnamefont{N.}~\bibnamefont{Yoshida}}, \bibnamefont{and}
  \bibinfo{author}{\bibfnamefont{N.}~\bibnamefont{Konno}},
  \bibinfo{journal}{Appl. Math. Comp.} \textbf{\bibinfo{volume}{126}},
  \bibinfo{pages}{255} (\bibinfo{year}{2002}).

\bibitem[{\citenamefont{Liggett}(1985)}]{liggett_85}
\bibinfo{author}{\bibfnamefont{T.~M.} \bibnamefont{Liggett}},
  \emph{\bibinfo{title}{Interacting Particle Systems}}
  (\bibinfo{publisher}{Springer-Verlag}, \bibinfo{address}{New York},
  \bibinfo{year}{1985}).

\bibitem[{\citenamefont{Grassberger}(1982)}]{grassberger_zpb82}
\bibinfo{author}{\bibfnamefont{P.}~\bibnamefont{Grassberger}},
  \bibinfo{journal}{Z. Phys. B} \textbf{\bibinfo{volume}{47}},
  \bibinfo{pages}{365} (\bibinfo{year}{1982}).

\bibitem[{\citenamefont{Jensen}(1993)}]{jensen_prl93}
\bibinfo{author}{\bibfnamefont{I.}~\bibnamefont{Jensen}},
  \bibinfo{journal}{Phys. Rev. Lett.} \textbf{\bibinfo{volume}{70}},
  \bibinfo{pages}{1465} (\bibinfo{year}{1993}).

\bibitem[{\citenamefont{Hinrichsen}(2000)}]{hinrichsen_ap00}
\bibinfo{author}{\bibfnamefont{H.}~\bibnamefont{Hinrichsen}},
  \bibinfo{journal}{Adv. Phys.} \textbf{\bibinfo{volume}{49}},
  \bibinfo{pages}{815} (\bibinfo{year}{2000}).

\bibitem[{\citenamefont{Szab{\'o}}(2005)}]{szabo_jpa05}
\bibinfo{author}{\bibfnamefont{G.}~\bibnamefont{Szab{\'o}}},
  \bibinfo{journal}{J. Phys. A: Math. Gen.} \textbf{\bibinfo{volume}{38}},
  \bibinfo{pages}{6689} (\bibinfo{year}{2005}).

\bibitem[{\citenamefont{Szab{\'o} and Sznaider}(2004)}]{szabo_pre04a}
\bibinfo{author}{\bibfnamefont{G.}~\bibnamefont{Szab{\'o}}} \bibnamefont{and}
  \bibinfo{author}{\bibfnamefont{G.~A.} \bibnamefont{Sznaider}},
  \bibinfo{journal}{Phys. Rev. E} \textbf{\bibinfo{volume}{69}},
  \bibinfo{pages}{031911} (\bibinfo{year}{2004}).

\bibitem[{\citenamefont{Szab{\'o} et~al.}(2007)\citenamefont{Szab{\'o},
  Szolnoki, and Szanider}}]{szabo_pre07}
\bibinfo{author}{\bibfnamefont{G.}~\bibnamefont{Szab{\'o}}},
  \bibinfo{author}{\bibfnamefont{A.}~\bibnamefont{Szolnoki}}, \bibnamefont{and}
  \bibinfo{author}{\bibfnamefont{G.~A.} \bibnamefont{Szanider}},
  \bibinfo{journal}{Phys. Rev. E} \textbf{\bibinfo{volume}{76}},
  \bibinfo{pages}{051921} (\bibinfo{year}{2007}).

\end{thebibliography}

\end{document}